\documentclass[prl,aps,twocolumn,preprintnumbers,showpacs,superscriptaddress]{revtex4}
\usepackage[colorlinks=true, pdfstartview=FitV, linkcolor=red, citecolor=blue, urlcolor=blue]{hyperref}

\usepackage{graphics,graphicx,epsfig,bm}
\usepackage{epstopdf}

\begin{document}

\title{A Processor Core Model for Quantum Computing}

\date{\today}

\author{Man-Hong Yung}
\email[email: ]{myung2@uiuc.edu}

\affiliation{Physics Department, University of Illinois at
Urbana-Champaign, Urbana IL 61801-3080, USA}

\author{Simon C. Benjamin}

\affiliation{Department of Materials, Oxford University, Oxford OX1
3PH, United Kingdom}

\affiliation{Centre for Quantum Computation, Department of Physics,
Oxford University OX1 3PU, United Kingdom}

\author{Sougato Bose}

\affiliation{Department of Physics and Astronomy, University
College London, Gower Street, London WC1E 6BT}


\begin{abstract}
We describe an architecture based on a processing `core'
where multiple qubits interact perpetually, and a separate
`store' where qubits exist in isolation. Computation consists of
single qubit operations, swaps between the store and the core, and
\emph{free} evolution of the core. This enables computation using physical systems where the entangling interactions are `always on'. Alternatively, for switchable systems
our model constitutes a prescription for optimizing many-qubit gates. We discuss implementations of the quantum Fourier
transform, Hamiltonian simulation, and quantum error correction.
\end{abstract}

\maketitle

Typically, schemes for solid state quantum computing involve an
array of qubits with some form of direct physical interaction
coupling nearby elements \cite{Kane98}. In order to implement a
specific algorithm, these schemes require the experimentalist to
dynamically control the magnitude of each qubit-qubit interaction
- effectively to be able to switch it `on' and `off'. A common
idea for achieving this is to somehow dynamically manipulate the
wavefunction overlap between a pair of neighboring qubits, while other nearby qubits are decoupled. This appears
feasible, but {\em highly} challenging. Moreover, even if a
switching mechanism can be implemented, frequent switching is
likely to increase the rate of dechoerence. A deeper objection is
that, by having the majority of a system's interactions `off' at a
given moment, we are failing to maximally exploit its
computational potential.

Recently ideas have emerged \cite{zhou,B&B} for computation in
systems where the interaction remains always on. However, these
proposals find ways to {\em effectively} pacify an interaction,
and therefore one can make the same objection that they are not
exploiting the full entangling power of the device. One class of system
that does make full use of a set of permanent
interactions is the mirror-inversion chain\
\cite{Christandl04,Albanese04,Karbach05,Clark04,Yung04,Shi04}. A
chain of spins, with suitably engineered coupling strengths, has
the property that a qubit placed on one end will later manifest at
the other - even though at intervening times it is distributed
over the chain. When more than one qubit is placed on the chain,
each will manifest at the complimentary site - but typically the
qubits will have aquired an entangling phase. It has been observed
\cite{Clark04} that this phase could in principle be employed to
create certain classes of entangled state, graph states, which are
the resource for one-way computation.

\begin{figure}[t]
\centering
\includegraphics[width=8cm]{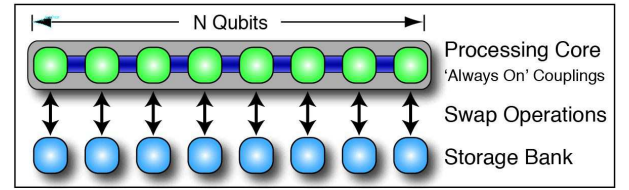}
\caption{ Architecture of the processor core model. The
core is an engineered spin-1/2 chain with always on
interactions. The storage bank consists of isolated sites where
qubits can be swapped to and from the corresponding sites in the
processing core. Controlled multi-target gates are constructed by
the free evolution of the spin chain.}\label{fig:arch}
\end{figure}

In this letter, we demonstrate the potential of such engineered spin chains
 to directly implement arbitrary controlled multi-qubit gates. The chain then
 acts as the computation core of our computer (see Figure \ref{fig:arch}) -  we need only supplement its free evolution
 with swap operations and single qubit manipulations. Note that this model is profoundly distinct from schemes involving a single qubit {\em bus}, e.g. the original ion trap schemes, since there the common mode represents only one qubit of information. We show that a controlled
multi-qubit gate can be constructed with exactly four free
evolutions of the spin chain, independent of the number of spins
involved. The controlling qubit can be any member of the spin chain, and the conditional
unitary operations applied to the target qubits can be of any type.
Such a gate can significantly reduce the number
of elementary operations for quantum algorithms involving many
non-local two-qubit operations.

We start with a finite chain of $N$ spin-${\textstyle{1 \over 2}}$
particles confied within local potentials and interaction with
their nearest neighbors. The Hamiltonian considered is:
\begin{equation}\label{eq:H}
H = \frac{1}{2}\sum_{j = 1}^{N-1} \omega_{j} \big( \sigma^{x}_{j}
\sigma^{x}_{j+1} + \sigma^{y}_{j} \sigma^{y}_{j+1} \big) +
\frac{1}{2} \sum_{j=1}^N \lambda_{j} \big(\sigma^{z}_j + 1\big) \, ,
\end{equation}
where the coupling constants $\left\{ {\omega _j ,\lambda _j }
\right\}$ are real and in general distinct. We will adopt the
convention that $\left| 0_j \right\rangle$ ($\left| 1_j
\right\rangle$) refers to be the spin-down state $\left|
\downarrow \right\rangle$ ($\left| \uparrow  \right\rangle$) at
the site~$j$.

Next, we require that the core should possess the so-called ``mirror
symmetry" \cite{Albanese04}, which implies the coupling constants
satisfy the relations $\omega _j  = \omega _{N-j}$ and $\lambda _j =
\lambda _{\bar j}$ where $\bar j \equiv N {-} j {+} 1$ denotes the
mirror-conjugate site of~$j$. Let $ \left| {s} \right\rangle \equiv
\left| {s_1 s_2 \cdots s_N } \right\rangle$, $s_j = \left\{ {0,1}
\right\}$, be a particular spin configuration. Mirror inversion is
said to occur when the state $\left| s \right\rangle$ is driven by
the evolution operator $U\left( \tau  \right) = e^{ - iH\tau }$, for
some fixed time period $\tau$, to the inverted state $\left| {\bar
s} \right\rangle \equiv \left| {s_N \cdots s_2 s_1 } \right\rangle$
(up to a phase factor). Note that the term {\em mirror inversion}
refers to the inversion of classical states $\left| s \right\rangle$
in which all the sites have definite spin values. For a quantum
state, being a superposition of the basis states in general, some
internal phases will be acquired. However, it is exactly these
entangling phases which we will exploit for gate construction.

It is shown \cite{Yung04} that the sufficient and necessary
condition for mirror inversion in mirror symmetrical chains is
determined by the eigenvalue spectrum of $H_S$ ($\hbar=1$), the
single excitation subspace of $H$. Then,
\begin{equation}\label{eq:condition}
e^{ - iE_k \tau }  = \left( { - 1} \right)^{k} e^{ - i\phi _N }
\quad ,
\end{equation}
where $E_k$, $k=0,1,2,\cdots,N-1$, is the $\left( {k + 1} \right)
$-th eigenvalue of $H_S$ and $\phi_N$ is some global phase
independent of $k$. Due to the mirror symmetry, the coupling
constants $\left\{ {\omega _j ,\lambda _j } \right\}$ can be
determined by the eigenvalue spectrum. It is therefore an inverse
eigenvalue problem \cite{Parlett98}. Recently, many spectra
\cite{Christandl04,Albanese04,Shi04} satisfying the condition in
Eq.(\ref{eq:condition}) have been proposed. However, to keep our
model general, we will continue our discussion without reference
to any specific type of spectrum.

To construct multi-qubit gates, we need to know the matrix elements
of the evolution operator $U(\tau)$ in the $\left| s \right\rangle$
basis. Let $\textsf{U} \equiv U\left( \tau \right)$. By mapping our
picture of localized spins to that of spinless fermions
\cite{Albanese04,{Yung04}}, one can show that
\begin{equation}\label{eq:U}
\textsf{U} \left| s \right\rangle  = e^{ - n i \phi _N } \left( { -
1} \right)^{\left( {n - m} \right)/2} \left| {\bar s} \right\rangle
\quad,
\end{equation}
where $n$ is the number of spin-up states in $\left| s
\right\rangle$, and $m{=}0~(1)$ if $n$ is even (odd). The factor
$\left( { - 1} \right)^{( {n - m})/2}$ could be understood
intuitively as follows: if n is even (i.e. $m{=}0$), then the
operation of mirror inversion (re-ordering the state) is equivalent
to swapping $n/2$ pairs of fermions. Similarly for odd n, except the
factor should be the same as that of $n{-}1$ fermions. The phase
factor $e^{ - n i \phi _N }$ is a consequence of
Eq.(\ref{eq:condition}). In fact, the phase $\phi_N$ can be set to
zero if an appropriate spectrum of $H_S$ is chosen. In this case, it
has been demonstrated \cite{Clark04} that the operator $U(\tau)$
alone can generate a fully connected graph state. However, in
constructing multi-qubit gates, spurious correlations among qubits
in the graph state have to be eliminated. This can be achieved with
the help of an ancilla qubit within the storage array, initialized
to be $\left| 0 \right\rangle _a$. Let $\textsf{S}_x$ represent the
swap operation between the spin at the site $\bar{x}$ (compliment of
$x$) and the ancilla. We apply $\textsf{S}_x$ to the state in
Eq.(\ref{eq:U}) and allow the engineered chain to evolve once more,
i.e. applying $\textsf{Z}^x \equiv \textsf{U} \textsf{S}_x
\textsf{U}$ \cite{notations} to the initial state $\left| 0
\right\rangle _a \otimes \left| s \right\rangle$. Then from
(\ref{eq:U}) the final state is
\begin{equation}\label{eq:Z_x}
e^{ - \left( {2n - s_x } \right)i\phi _N } \left( { - 1}
\right)^{s_x \left( {n - 1} \right)} \left| {s_x } \right\rangle _a
\otimes \left| {s_1 s_2  \cdots 0_x  \cdots s_N } \right\rangle  \,
,
\end{equation}
which is the same for $n$ being either odd or even. Here we have
only assumed a swap operation performed between the site $\bar{x}$
and the ancilla. Therefore, the qubit staying at the ancilla spin
cannot be transferred back to the spin chain at this stage.
However, as we shall see (cf. Eq.(\ref{eq:W^x})), a more general
multi-qubit gate can be constructed based on $\textsf{Z}^x$ and
all of the qubits can reside in their original locations at the end of the
operation.

The next step is to interpret the result  (\ref{eq:Z_x}) in
terms of the quantum circuit model. The phase factor $e^{ - \left(
{2n - s_x } \right)i\phi _N }$ can be regarded as a result of $N$
local phase gates $\textsf{R}_j\left(-2\phi_N\right)$, where
$\textsf{R}_j\left(\varphi \right) \equiv \left| 0_j \right\rangle
\left\langle 0_j \right| + e^{  i\varphi} \left| 1_j \right\rangle
\left\langle 1_j \right|$, acting on all qubits, and one extra
phase gate, $\textsf{R}_j(\phi_N)$ acting on the spin at site $x$
along.
On the other hand, the factor $\left( { - 1} \right)^{s_x \left(
{n - 1} \right)}$ can be considered as due to the application of
controlled-$\sigma_z$ to all qubits, except the spin at site $x$
which is encoded with the controlling qubit. Suppose we now apply
local operations to get rid all of the controlled phase gates
$\textsf{R}_j$ (or simply choose an eigenvalue spectrum such that
$\phi_N=0$), effectively we have constructed a controlled
multi-target gate, which requires two free evolutions of the
engineered Hamiltonian for any $N$. Note that for this multi-qubit
gate generated by $\textsf{Z}^x$, the $\sigma_z$ gate has to be
applied to \emph{all} qubits, controlled by a single qubit at site
$\bar{x}$. However, the $\sigma_z$ gate can be converted into
controlled-$\textsf{V}_j$ \cite{Nielsen_book}, where
\begin{equation}
\textsf{V}_j  = \left( {\begin{array}{*{20}c}
   {\sin \theta _j } & {e^{i\varphi _j } \cos \theta _j }  \\
   {e^{ - i\varphi _j } \cos \theta _j } & { - \sin \theta _j }  \\
\end{array}} \right) \quad ,
\end{equation}
through local operations $\textsf{A}_j$ provided that the relations
$\textsf{A}_j \, \sigma_z \textsf{A}^\dag_j = \textsf{V}_j$ and
$\textsf{A}_j \textsf{A}^\dag_j = \textsf{I}_j$, where
$\textsf{I}_j$ is the identity operator, are satisfied. For example,
the case $\theta_j=\varphi_j {=} 0$ corresponds to a $\sigma_x$
gate. Operationally, we denote the construction of this controlled
multi-target gate by $\textsf{V}^x \equiv \textsf{A} \textsf{Z}^x
\textsf{A}^\dag$, where $\textsf{A} \equiv \prod\nolimits_{j =1 }^N
{\textsf{A}_j }$ (and similarly for $\textsf{A}^\dag$). An immediate
application of this gate is that, if we initialize the controlling
qubit to be $\left| 0 \right\rangle + \left| 1 \right\rangle$ and
the rest $\left| {000...0} \right\rangle$, it can efficiently
generate a cat state $\left| {000...0} \right\rangle  + \left|
{111...1} \right\rangle$, which is interesting for various
applications including single qubit measurement and encoding error
correcting codes, such as the Shor's code.

\begin{figure}[t]
\centering
\includegraphics[width=7cm]{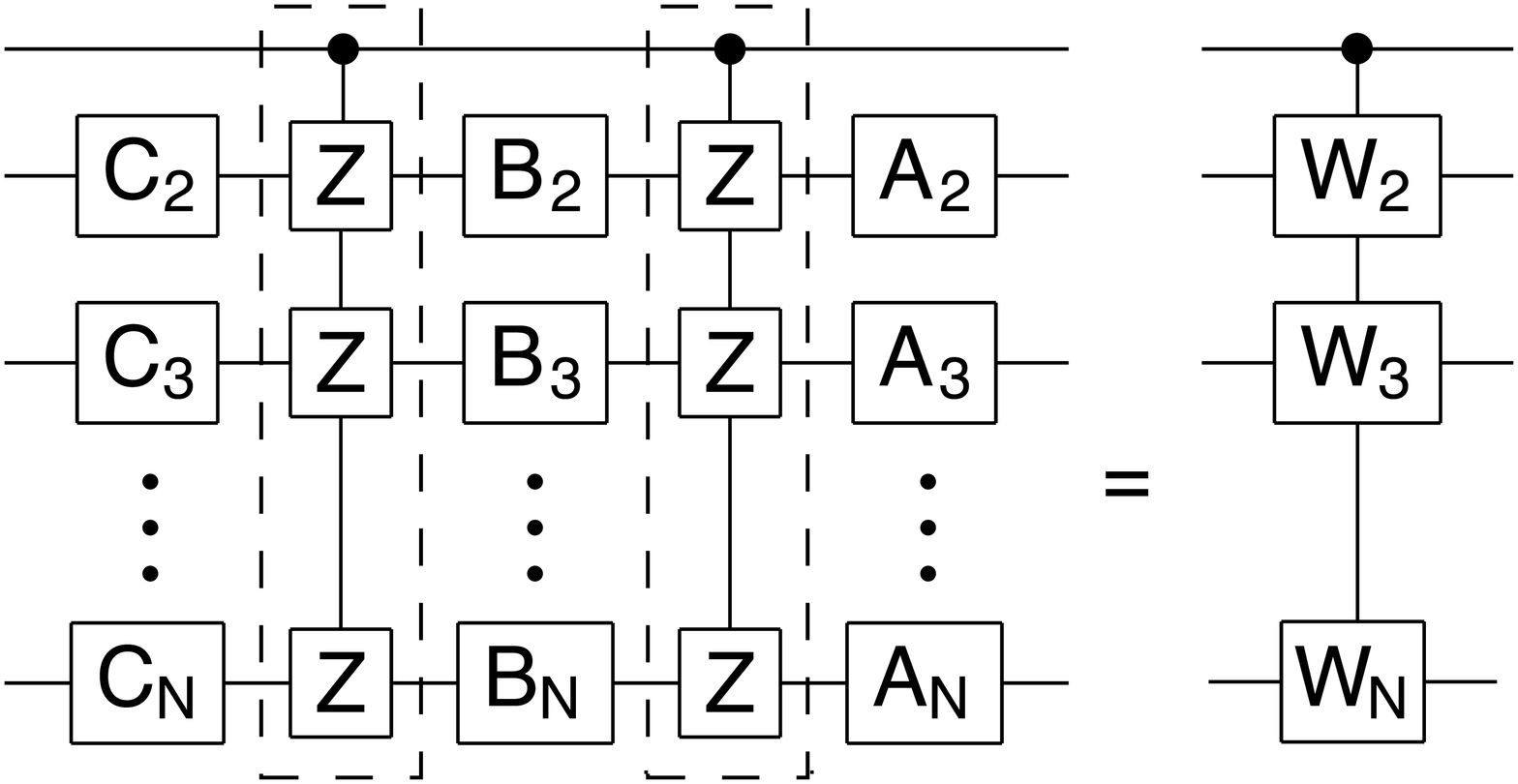}
\caption{The circuit diagram of a controlled multi-target
gate~$\textsf{W}^1=\textsf{A} \textsf{Z}^1 \textsf{B} \textsf{Z}^1
\textsf{C}$ constructed by the free evolution of an engineered spin
chain `core'. The parts inside the dashed boxes are $\textsf{Z}^1$ which
involves two free evolutions of the spin chain and one ancilla in the `store'. The
local operators $\textsf{A}_j$, $\textsf{B}_j$ and $\textsf{C}_j$
satisfy $\textsf{A}_j \textsf{B}_j \textsf{C}_j =\textsf{I}_j$ and
$\textsf{A}_j \textsf{Z} \textsf{B}_j \textsf{Z} \textsf{C}_j =
\textsf{W}_j $, where the unitary operations $\textsf{W}_j$, up to a
phase, are defined in (\ref{eq:W_j}). }\label{fig:gate}
\end{figure}

The controlled operations $\textsf{V}_j$ are not yet completely general: for example, the
phase gate $\textsf{R}_j(\varphi)$ and the identity
operator~$\textsf{I}_j$ are excluded.
We can construct a more general controlled multi-target
gate, which applies arbitrary unitary operations $\textsf{W}_j$ on
the qubits. To proceed, consider applying $\textsf{U}$ to the
state $\left| {s_1 s_2  \cdots 0_x \cdots s_N } \right\rangle$,
which is assumed to contain $n$ spin-up states. The phase factors
generated are exactly the same as that in Eq.(\ref{eq:U}). Now we
can apply $\textsf{U} \textsf{S}_x$ to the resulting state;  $\textsf{S}_x$ returns $\left| {s_x } \right\rangle_a$ from the ancilla to the core. The final state is $e^{ - \left( {2n + s_x }
\right)i\phi _N } \left( { - 1} \right)^{s_x n} \left| s
\right\rangle$, which can also be considered as an controlled
multi-target gate analogous to the one generated by $\textsf{Z}^x
\left| s \right\rangle$. For simplicity, we again assume
$\phi_N=0$. To construct a more general unitary matrix
\begin{equation}\label{eq:W_j}
\textsf{W}_j  = \left( {\begin{array}{*{20}c}
   {e^{i\left( {\alpha _j  - \beta _j  - \delta _j } \right)} \cos \gamma _j } & { - e^{i\left( {\alpha _j  - \beta _j  + \delta _j } \right)} \sin \gamma _j }  \\
   {e^{i\left( {\alpha _j  + \beta _j  - \delta _j } \right)} \sin \gamma _j } & {e^{i\left( {\alpha _j  + \beta _j  + \delta _j } \right)} \cos \gamma _j }  \\
\end{array}} \right) \, ,
\end{equation}
including the identity operator, one can always choose a set of
local operations $\textsf{A}_j$, $\textsf{B}_j$ and $\textsf{C}_j$
such that the relations $\textsf{A}_j \textsf{B}_j \textsf{C}_j =
\textsf{I}_j$ and $ e^{i\alpha _j } \textsf{A}_j \textsf{Z}
\textsf{B}_j \textsf{Z} \textsf{C}_j = \textsf{W}_j$
\cite{Nielsen_book} are satisfied. Our result is then that the
controlled-$W_j$ gate $\textsf{W}^x$, with controlling qubit at
site $x$, can be constructed by the following sequence of
operations: $\textsf{W}^x = \textsf{A} \textsf{Z}^x \textsf{B}
\textsf{Z}^x \textsf{C}$, where $\textsf{A}$, $\textsf{B}$ and
$\textsf{C}$ are the tensor products of local operators
$\textsf{A}_j$, $\textsf{B}_j$ and $\textsf{C}_j$ respectively.
Equivalently,
\begin{equation}\label{eq:W^x}
\textsf{W}^x  =  \left| {0_x } \right\rangle \left\langle {0_x }
\right|  \otimes \textsf{I} +   \left| {1_x } \right\rangle
\left\langle {1_x } \right| \otimes \prod\limits_{j \ne x}
{\textsf{W}_j } \quad .
\end{equation}
The circuit diagram for an example $\textsf{W}^1$ is shown in
Figure~\ref{fig:gate}. For any $N$, the cost of generating
$\textsf{W}^x$ includes four free evolutions of the engineered
chain, two swap operations with the same ancilla and local
operations.

We now describe the application of this model to the key problems of quantum Fourier transform, Hamiltonian simulation and quantum error correction. When we wish to make a statement about
the efficiency of our {\em processor core} model, we will compare it with a notional {\em fully-switched} system having a Hamiltonian similar to (\ref{eq:H}), except that the interactions $\omega_i$ can be independently switched on and off. We assume the fully-switched system couples only pairs of qubits simultaneously, although it may do so in parallel - i.e. $\omega_i\omega_{i+1}=0$ at all times. A real fully-switched system could, presumably, activate several adjacent interactions: the results we describe here can be seen as a prescription for doing precisely that, in order to increase efficiency. The primary gain in efficiency will of course be a reduction in the number of switching events (and consequently, a potential reduction in the decoherence rate) - but remarkably there can also be an absolute speed-up by a fixed factor, as we presently discuss.

One of the immediate applications of the controlled multi-target
gate $\textsf{W}^x$ is the operation of quantum Fourier transform
(\textsf{QFT}), which is a key ingredient in many quantum algorithms
such as the Shor's algorithm. In the above notations, the standard
\textsf{QFT} circuit can be constructed by applying the multi-qubit
gates and the Hadamard gates alternatively,
\begin{equation}
\textsf{QFT} = \textsf{H}_N \textsf{W}^{N - 1} \textsf{H}_{N - 1}
\cdots \textsf{W}^2 \textsf{H}_2 \textsf{W}^1 \textsf{H}_1 \quad ,
\end{equation}
where $\textsf{W}_j  = \textsf{R}_j \left( {\pi /2^{j - x} }
\right)$ for $j>x$ and $\textsf{W}_j=\textsf{I}_j$ otherwise. Here
each joint operation $\textsf{W}_j$ costs exactly four free
evolutions, including two swaps. The QFT circuit depth is
therefore $O(N)$, as is the total number of switching events. For
the switched model, the circuit depth is also $O(N)$, while the
absolute number of switching events is $O(N^2)$.

The second application is the simulation of the evolution of an
``artificial" Hamiltonian $H_A$ formally representing a joint
interaction between $r$ spin-1/2 particles,
\begin{equation}\label{eq:H_A}
H_A  = \sigma _1^z  \otimes \sigma _2^z  \otimes \sigma _3^z \cdots
\otimes \sigma _r^z \quad ,
\end{equation}
which is locally equivalent to the class of the Hamiltonian of the
form $\sigma _1^{w_1 }  \otimes \sigma _2^{w_2 }  \otimes \sigma
_3^{w_3 }  \cdots  \otimes \sigma _r^{w_r }$, where $\sigma
_j^{w_j } = \sigma _j^x , \, \sigma _j^y$ or $\sigma _j^z$.
Although it is unlikely to find a group of spin-1/2 particles
interacting naturally under the Hamiltonian $H_A$, some higher
dimensional systems can be mapped by those two-level systems.
Moreover, the form of $H_A$ can be considered as basic building
blocks for simulating more complex Hamiltonians through local
operations and the approximation: $e^{i\left( {A + B}
\right)\Delta t} = e^{iA\Delta t} e^{iB\Delta t} + O\left( {\Delta
t^2 } \right)$ for short time-interval~$\Delta t$.

Consider an engineered core of $N+1$ spins, with $\phi_N=0$,
initialized as $\left| {0s} \right\rangle  \equiv \left| {0s_1 s_2
...s_N } \right\rangle$. The quantum circuit for simulating the
evolution operator $U_A \left( {\Delta t} \right) = e^{ - iH_A
\Delta t}$ for (\ref{eq:H_A}) can be constructed by the following
sequence of operations \cite{Nielsen_book}:
\begin{equation}
U_A \left( {\Delta t} \right) = \textsf{H}_0 \textsf{W}^0
\textsf{H}_0 \textsf{T}_0\left( {\Delta t} \right) \textsf{H}_0
\textsf{W}^0 \textsf{H}_0 \quad ,
\end{equation}
where $\textsf{T}_0 \left( {\Delta t} \right) = \exp \left( { -
i\sigma^z_0 \Delta t} \right)$ and $W_j  = \sigma _j^z$ (or $W_j
= I_j$ if the qubit at site $j$ is not involved). The basic idea
of this construction is to store the parity (i.e. $m=\{0,1\}$) of
the sites $j=1,2,3,\ldots,N$ to site $0$. The phase generated by
the local operation $\textsf{T}_0$, depending on the parity, is
exactly the one required for $H_A$. From (\ref{eq:W^x}), it is
apparent that the series of non-local operations $\textsf{W}^0$
can be achieved by four free evolutions of the engineered chain of
$N\geq r$ spins, with the aid of an ancilla and local operations.
This scheme offers a flexibility of generating interactions
involving various number of spins using the same spin chain. The
costs of generating each type of interactions are fixed (eight
free evolutions). Alternatively, if one just needs to generate
interactions with \emph{fixed} number of qubits, i.e. $r=N$, the
cost can be reduced to two free evolutions and no ancilla is
needed. The sequence of operations in this case is
\begin{equation}\label{eq:H_A2}
U_A \left( {\Delta t} \right) = \textsf{H}_0 \textsf{U}
\textsf{H}_{\bar 0} \textsf{T}_{\bar 0} \left( {\Delta t} \right)
\textsf{H}_{\bar 0} \textsf{U} \textsf{H}_0 \quad .
\end{equation}
The crucial observation for obtaining (\ref{eq:H_A2}) is that
$\textsf{H}_{\bar 0} \textsf{U} \textsf{H}_0 \left| {0s}
\right\rangle  = \left( { - 1} \right)^{\left( {n - m} \right)/2}
\left| {\bar sm} \right\rangle$. Thus, the desired phase can be
obtained by applying the local operator $\textsf{T}_{\bar 0}$ at
site~$\bar 0$.

The processor core model will also be advantageous in running
quantum algorithms in a fault tolerant fashion with concatenated
code-words. For example, in the Steane code, six gates of the
class $\textsf{W}^x$ are required for error syndrome
  measurement (Fig.10.16 of Ref.\cite{Nielsen_book}). In our approach, each
  level of concatenation just multiplies the number of targets in
  each $\textsf{Z}^x$ by $7$ \cite{note-on-concat} but does not increase the number of applications of such gates. However,
  the number of elementary switching operations required by a fully-switched system to realize $\textsf{W}^x$ increases
  $7$ fold with {\em each} level of concatenation.

The discussions above highlight potential gains in terms of
simplicity: the circuit depth, or total number of switching
operations. It is also interesting to ask, can the total time
required be reduced by applying the processor core model? We can
quickly conclude that any speed-up must be bounded, since the
fundamental operation $U$ (eqn. \ref{eq:U}) can be simulated on a
fully-switched array in time $O(N)$\ \cite{simulateU}, while $U$
also takes time $O(N)$ to evolve on our processor-core (given a
fixed maximum interaction strength).
Interestingly, there can be speed-ups within this constraint, i.e.
by bounded factors. To make a definite statement we specialize to a
core with a linear spectrum, i.e. $\Delta_k = E_k {-} E_{k+1}$ being
constant (e.g. \cite{Christandl04,{Albanese04},{Shi04}} ) since this
is the time-optimal choice for a given spectral {\em range}\
\cite{range}. Let us the compare the time required for a
\emph{simple state transfer}, i.e. $\left| {100...0} \right\rangle
\to \left| {0...001} \right\rangle $, from one end of the chain to
the other. If we now assume for simplicity that that the maximum
interaction strength $\omega_{\max}{\simeq} N/4$ (see Ref.
\cite{Christandl04}) scales as $N$, then the time required for the
processor-core evolution is simply $\pi$ for all $N$. On the
fully-switched system, the time required for each swap is
${\textstyle{\pi  \over {2\omega _j }}}$, thus the total time
required is $T\left( N \right) = \sum\nolimits_{j = 1}^{N - 1}
{{\textstyle{\pi  \over {2\omega _j }}}}$. (Recall that our
fully-switched system can couple only pairs of qubits
simultaneously, thus each state transfer must complete before the
next is initiated). One can easily show that $ T\left( N \right) \ge
\left( {N {-} 1} \right)\pi /2\omega _{\max }  \simeq 2\pi \left( {N
{-} 1} \right)/N$. Thus this {\em always-on} processing core can be
superior by a factor of $2$ for large $N$.



Finally, we remark that the periods of free processor core evolution
can be relatively robust versus timing errors $\tau \to \tau +
\delta t$ in subsequent swaps to the store. Consider the most
general case, we start with the state $\left| {\bar \psi }
\right\rangle  = \sum\nolimits_j {\alpha _j \left| {\bar s_j }
\right\rangle }$, where $\sum\nolimits_j {} \left| {\alpha _j }
\right|^2 {=} 1$. If the evolution time is taken perfectly, we
expect the final state to be $\left| \psi \right\rangle  =
\sum\nolimits_j {\alpha _j e^{i\phi _j } \left| {s_j } \right\rangle
}$, where $\phi _j $ represents the overall phase in Eq.
(\ref{eq:U}) for the spin configuration ${\left| {s_j }
\right\rangle }$. If not, we have $\left\langle \psi \right|U\left(
{\tau  {+} \delta t} \right)\left| {\bar \psi } \right\rangle  =
\sum\nolimits_{j,k} {\alpha _j^* \alpha _k e^{i\left( {\phi _k  -
\phi _j } \right)} \left\langle {s_j } \right|U\left( {\delta t}
\right)\left| {s_k } \right\rangle } \equiv 1 + iA\delta t + B\delta
t^2  + O\left( {\delta t^3 } \right)$. Here both $A$ and $B$ are
real. The error, defined as $ \epsilon  \equiv 1 - \left|
{\left\langle \psi  \right|U\left( {\tau {+} \delta t} \right)\left|
{\bar \psi } \right\rangle } \right|^2 $, is therefore just second
order in $\delta t$.


In conclusion, we have demonstrated how to construct controlled
multi-target gates through the natural evolution of a processor
core where interactions are {\em always on}. This model allows
computation with physical systems where the entangling interactions are not
switchable. Alternatively, in switchable systems our protocol can
play an important role in simplifying
multi-qubit operations. We demonstrated this by showing that the
{\em fully-switched} model is fundamentally more complex for certain important
algorithmic tasks. For various operations spanning many qubits, including long range state transfer, the temporal
requirement for the processor core model is less-than-or-equal-to that of the fully-switched model. Therefore, in terms of the
reduction of dynamical control while maintaining the same {\em speed},
many schemes presently described in terms of two-qubit gates can be enhanced by
incorporating the processor core concept.



\begin{acknowledgments}
MHY acknowledges the support of the Croucher Foundation and the QIPIRC. SCB is supported by the Royal Society.
\end{acknowledgments}




\begin{thebibliography}{99}
\bibitem{notations}
Subscripts denote the site to be acted upon and superscripts denote
the site of the controlling qubit.
\bibitem{simulateU}
One can simulate $\textsf{U}$ on a 1D full-switched system using parallel swap operations performed over $N$ time steps.
One simply alternates between swaps on all pairs $(s_{2i},s_{2i+1})$, and then all pairs $(s_{2i-1},s_{2i})$. Each qubit will, at some point, be adjacent to
every other, thus by interleaving nearest-neighbor phase gates with
the swap operations, we implement the correct phase, still within
$O(N)$ time steps. The total number of gates is $O(N^2)$.
\bibitem{range} According to Eq.~(\ref{eq:condition}) the phase
difference $\Delta_k \tau$ must be at least $\pi$ for
$k=0,1,2,...,N{-}2$. To compare different eigenvalue spectra, we
require the range to be normalized $E_{\max}{-}E_{\min}=1$. The
linear spectrum requires the minimal evolution time to be $\tau =
(N{-}1)\pi$. For any eigenvalue spectrum which is non-linear, there
must exist at least one $\Delta_k < 1/(N{-}1)$, and hence $\tau >
(N{-}1)\pi$.
\end{thebibliography}
\end{document}